\documentclass[journal=cmatex]{achemso}
\usepackage[T1]{fontenc} 
\usepackage[version=3]{mhchem}
\usepackage{amsmath}
\usepackage{amssymb}
\usepackage{graphicx}
\usepackage{subcaption}
\usepackage{color}
\usepackage{dcolumn}
\usepackage{bm}
\usepackage{miller}
\usepackage{pdfpages}
\usepackage[utf8]{inputenc}
\usepackage{hyperref}

\newcommand*{\rtten}[1]{\mathbf{\boldsymbol{#1}}}

\graphicspath{ {Figures/} }
\author{Zeeshan Ahmad}
\affiliation{Department of Mechanical Engineering, Carnegie Mellon University, Pittsburgh, Pennsylvania 15213, USA}
\author{Victor Venturi}
\affiliation{Department of Mechanical Engineering, Carnegie Mellon University, Pittsburgh, Pennsylvania 15213, USA}
\author{Hasnain Hafiz}
\affiliation{Department of Mechanical Engineering, Carnegie Mellon University, Pittsburgh, Pennsylvania 15213, USA}
\author{Venkatasubramanian Viswanathan}
\affiliation{Department of Mechanical Engineering, Carnegie Mellon University, Pittsburgh, Pennsylvania 15213, USA}
\alsoaffiliation[phys]{Department of Physics, Carnegie Mellon University, Pittsburgh, Pennsylvania 15213, USA}
\email{venkvis@cmu.edu}

\title[ionic-cond]
  {Interfacial Effects on Solid Electrolyte Interphase in Lithium-ion Batteries}

\begin{document}


\begin{tocentry}
\includegraphics[scale=0.45]{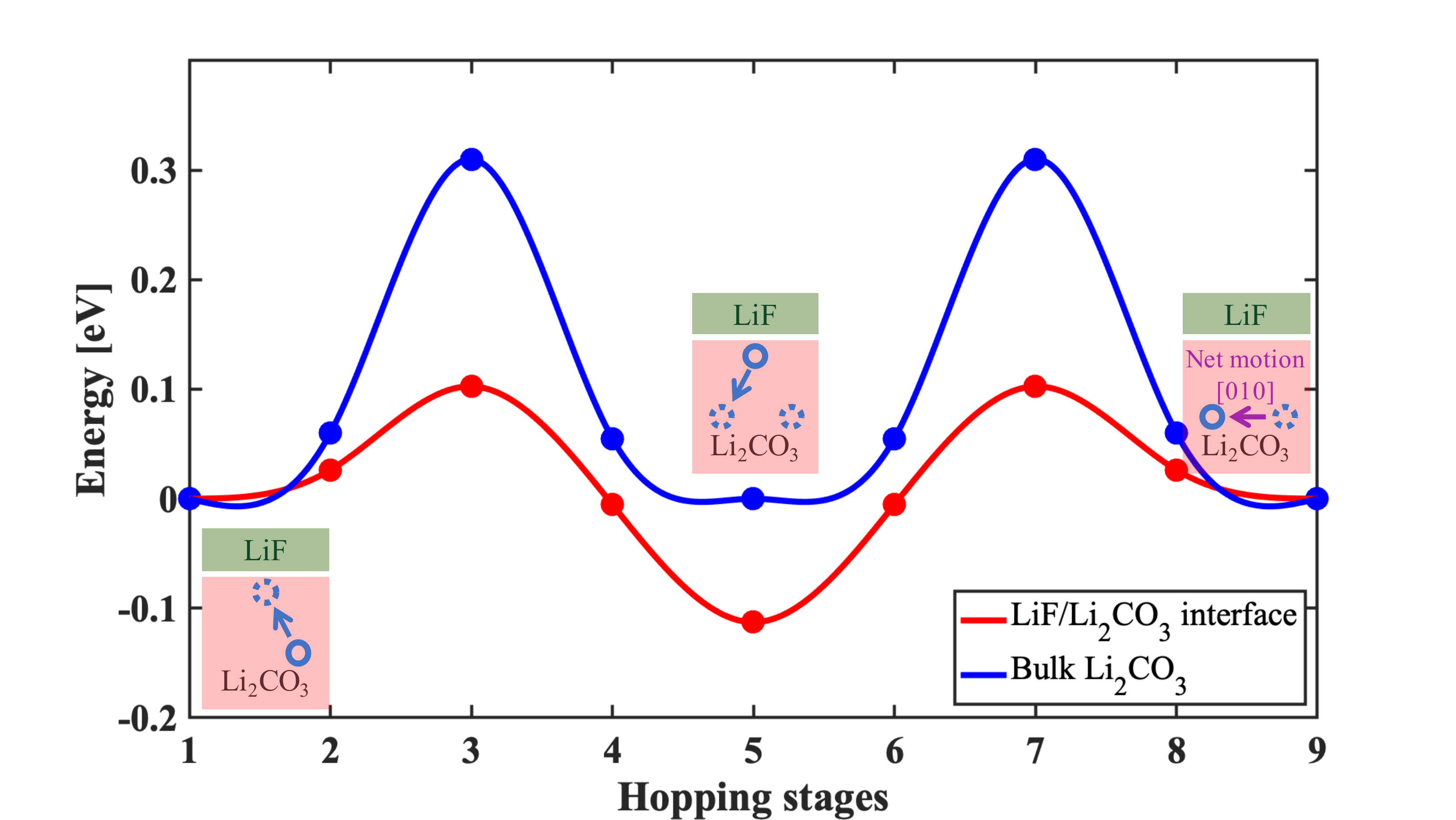}
\end{tocentry}

\begin{abstract}
The existence of passivating layers at the interfaces is a major factor enabling modern lithium-ion (Li-ion) batteries. Their properties determine the cycle life, performance, and safety of batteries. A special case is the solid electrolyte interphase (SEI), a heterogeneous multi-component film formed due to the instability and subsequent decomposition of the electrolyte at the surface of the anode. The SEI acts as a passivating layer that hinders further electrolyte disintegration, which is detrimental to the  Coulombic efficiency. In this work, we use first-principles simulations to investigate the kinetic and electronic properties of the interface between lithium fluoride (LiF) and lithium carbonate (\ce{Li2CO3}), two common SEI components present in Li-ion batteries with organic liquid electrolytes. We find a coherent interface between these components that restricts the strain in each of them to below 3\%. We find that the interface causes a large increase in the formation energy of the Frenkel defect, generating Li vacancies in LiF and Li interstitials in \ce{Li2CO3} responsible for transport. On the other hand, the Li interstitial hopping barrier is reduced from $0.3$ eV in bulk \ce{Li2CO3} to $0.10$ or $0.22$ eV in the interfacial structure considered, demonstrating the favorable role of the interface. Controlling these two effects in a heterogeneous SEI is crucial for maintaining fast ion transport in the SEI. We further perform Car-Parrinello molecular dynamics simulations to explore Li ion conduction in our interfacial structure, which reveal an enhanced Li ion diffusion in the vicinity of the interface.  Understanding the interfacial properties of the multiphase SEI represents an important frontier to enable next-generation batteries.
\end{abstract}


\section{Introduction}
Lithium-ion (Li-ion) batteries have revolutionized consumer electronics and road transportation~\cite{li201830, blomgren2016development} and offer a promising route to achieve electric aviation~\cite{Bills2020performance, kuhn2011renewable, krishnamurthy2020beyond}. The interface between the electrode and the electrolyte in a battery, and the related interphases formed, strongly influence the Coulombic efficiency, cycling performance, and power capability of the battery~\cite{Banerjee2020interfaces, Wang2018review}. The properties of a surface or an interface can be vastly different from those of the bulk material. For example, lithium iodide doped with alumina has a Li-ion conductivity orders of magnitude higher than any of the individual components~\cite{Liang1973conduction} due to the space charge effect. Most solid and liquid electrolytes are not electrochemically or chemically stable at the potentials encountered at the interfaces in Li metal batteries~\cite{Xu2004nonacqueous}. An ultrathin multicomponent interphase, commonly referred to as solid electrolyte interphase (SEI), first introduced by ~\citeauthor{Peled1979electrochemical}~\cite{Peled1979electrochemical}, is frequently generated as a result of both liquid and solid electrolyte decomposition in Li-ion batteries.  \citeauthor{Zhu2015origin}~\cite{Zhu2015origin} studied the voltage stability windows of some common inorganic solid electrolytes and found that none of them are thermodynamically stable against Li metal anode and form compounds such as \ce{LiF}, \ce{Li2O}, \ce{Li2S}, \ce{Li3P} and \ce{Li3N} at the low voltages encountered at anode. 
The SEI plays a major role in kinetically stabilizing the interface and preventing continued electrolyte decomposition, which can lower the Coulombic efficiency. Several studies have demonstrated a drastic improvement in battery performance by engineering the SEI, including enabling high energy density Li-metal anodes~\cite{Jin2017self-healing, liu2017artificial, Suo2015water, Cheng2015review, Zhu2020design}.

With organic liquid electrolytes, the SEI is composed of an inner dense (closer to anode) and an outer porous layer~\cite{verma2010review}. The passivating film is formed immediately as the anode is placed in contact with the electrolyte. The inner dense layer is composed of crystalline components such as \ce{Li2CO3}, \ce{LiF} and \ce{Li2O} while the outer layer is composed of organic materials~\cite{Xu2004nonacqueous}.  Recent direct operando observations have also confirmed the existence of an inorganic SEI composed mainly of LiF and \ce{Li2CO3}~\cite{Mozhzhukhina2020direct}.  Typically, it is desired that  the SEI should allow the movement of Li-ions but be electronically insulating. Its thickness is determined by the electron tunneling range beyond which electrolyte decomposition stops~\cite{balbuena2004lithium}. As Li ion transport through the SEI is often the rate limiting step~\cite{jow2018factors}, understanding ion transport through it is crucial for improving the power capability of the battery. Furthermore, the uniform current density necessary throughout the anode-SEI interface for smooth electrodeposition in Li-metal batteries requires the existence of an SEI layer with homogeneous properties. Several modeling studies have explored the role of interfaces between electrolytes/SEI and the electrode in determining stability, kinetics, charge transfer, and wettability in Li-ion batteries~\cite{Leung2010abinitio,Li2016computational,Kim2019predicting,Leung2018kinetics,Ramasubramanian2020stability, Leung2020DFT}. The interfaces within the components of the mosaic SEI, on the other hand, may also hinder or enhance the ion transport in a Li-ion battery. ~\citeauthor{Zhang2016synergetic}~\cite{Zhang2016synergetic} proposed the synergistic effect between \ce{LiF} and \ce{Li2CO3} as the cause of efficient ion conduction in the SEI. They argued that the presence of two phases, \ce{LiF} and \ce{Li2CO3} promotes ion transport in \ce{Li2CO3} through the formation of interstitial Li in \ce{Li2CO3}. This was backed by calculations of defect formation energies in the bulk phases of LiF and \ce{Li2CO3}. However, new questions arise about the role of interfaces in promoting formation of defects and ion transport. Various mechanisms of enhancement of ion conduction due to space charge effects have been proposed~\cite{Maier1995ionic,Sata2000mesoscopic} and exploring the role of heterointerfaces through first-principles might shed more light on intricate physics of ion transport.

In this work, we perform an in-depth study of the effect of interfaces on Li ion hopping, defect formation, electronic transport, and dynamics by explicitly modeling interfaces between the two major components of the inner SEI layer: LiF and \ce{Li2CO3}. We find that the interface causes an increase in the formation energy of the defects, namely Li vacancies in LiF and Li interstitials in \ce{Li2CO3} that are responsible for ion transport in the SEI. The activation energy for Li ion hopping in \ce{Li2CO3} based on the knockoff mechanism is lowered by the interface. Molecular dynamics simulations are used to calculate the directional activation energies, showing the lowest activation energy of Li ion hopping along [010] direction of \ce{Li2CO3}.

\section{Computational Methodology}
\subsection{Density Functional Theory calculations}
Density functional theory (DFT) calculations were performed using Quantum Espresso plane wave \texttt{pw.x} code~\cite{Giannozzi2009quantum,Giannozzi2017advanced}. The Perdew–Burke-Ernzerhof (PBE) exchange-correlation functional was used for all DFT calculations. Energy cutoffs were at least 40 Ry for the wavefunction and 320 Ry for the electron density. Ultrasoft pseudopotentials from the GBRV library~\cite{Garrity2014pseudopotentials} were used for Li, C, O and F. These calculations were used to optimize the structures (lattices and ionic positions) of \ce{LiF} and \ce{Li2CO3} as well as the interface between the two for use in molecular dynamics simulations. These calculations were also used to obtain the defect formation energies.

For the calculation of hopping barriers, we used the nudged elastic band (NEB) method~\cite{berne1998classical} as implemented in the  projector augmented wave (PAW) package GPAW~\cite{enkovaara2010electronic}. Ionic relaxation was performed with  a force convergence criterion $<0.05$ eV/\AA. The atomic simulation environment~\cite{HjorthLarsen2017atomic} was used for structure manipulation along with the interface to GPAW and the VESTA package~\cite{Momma2011vesta} was used for visualization.

\subsection{Molecular Dynamics simulations} 
First-principles Car-Parrinello molecular dynamics~\cite{Car1985unified} simulations were performed using the \texttt{cp.x} code in Quantum Espresso~\cite{Giannozzi2009quantum, Giannozzi2017advanced}. We used a time step of 6 au (0.145 fs) with an effective electron mass of 500 au for the electron. The kinetic energy cutoffs used for wavefunction and charge density used were 30 Ry and 300 Ry respectively. The molecular dynamics simulations were performed at temperatures of 600, 800 and 900 K. The DFT-optimized interface was used as the initial structure for the molecular dynamics simulations. A random displacement was then introduced in the ionic positions. The conjugate gradient electron dynamics was used initially and then periodically to bring the electrons on the Born-Oppenheimer surface.  The temperature was increased to the required value by setting up a Nose-Hoover thermostat chain~\cite{Nos1984unified, Nos1984molecular, Hoover1985canonical} over a time of $\sim$7 ps. The production runs involving diffusivity calculations were done using the microcanonical ensemble to prevent the thermostat from affecting the trajectory~\cite{Marcolongo2017ionic}. In this work, we only consider the diffusivity of Li atoms in \ce{Li2CO3} since \ce{Li2CO3} is contains relatively more mobile Li ions (activation energy of hopping much lower) and is primarily responsible for conduction. This approach allows us to circumvent any effects that the low number of LiF layers used in the simulation may have on our final results.

\section{Results and Discussion}

\subsection{Structure of the Interface}
LiF exists in the rocksalt structure (space group 225, Fm$\bar{3}$m) with a mixture of edge and corner sharing \ce{LiF6} octahedra. From DFT calculations, we obtain its lattice constant $a=4.0657$~{\AA}, which agrees well with the experimental value of 4.0173 \AA ~\cite{Wyckoff1963cryst}.
\ce{Li2CO3} crystallizes in the monoclinic structure with space group 15, C2/c with planar \ce{CO3$^{2-}$} groups separated by \ce{Li+} ions. Our computed lattice parameters for \ce{Li2CO3} are $a=8.369$ {\AA}, $b=5.013$ \AA, $c=6.345$ \AA, $\alpha, \gamma = 90^\circ$ and $\beta=114.093^\circ$ in close agreement with experimental values~\cite{Dunstan2015ion,Idemoto1998crystal}.

A unique interface between \ce{LiF} and \ce{Li2CO3} is identified by the Miller indices of the surfaces of the two materials facing the interface and the two lattice vectors along the interface. We generate a coherent interface by aligning the lattice vectors of LiF and \ce{Li2CO3} with angle $90^{\circ}$ between them and repeating the in plane lattice vectors. The lowest index (001) surface of LiF is placed parallel to the plane of the interface. Among the low index surfaces, this leaves the choice of (001) or (100) surface of \ce{Li2CO3} in plane of the interface. Of these, the (001) surface resulted in lower strain/smaller lattice matched unit cell and was chosen for the simulations. The $c$ axis of \ce{Li2CO3} is perpendicular to the interface. Considerations of coherency strain minimization and computational tractability led to the following multiplicities of the conventional unit cell lattice vectors: $(1, 4, 2)$ for \ce{Li2CO3} and $(2,5,1)$ for \ce{LiF}. These supercells resulted in a coherency strain of less than 3 \% for LiF and less than 2 \% for \ce{Li2CO3}. Li, O or C terminations are possible for the  \ce{Li2CO3} (001) surface, out of which we chose the Li termination to prevent breaking the C-O bonds. This termination breaks one Li-O bond per Li and reduces the number of Li-O bonds for the surface Li atoms from 4 to 3.

\begin{figure}
    \centering
    \includegraphics[scale=0.8]{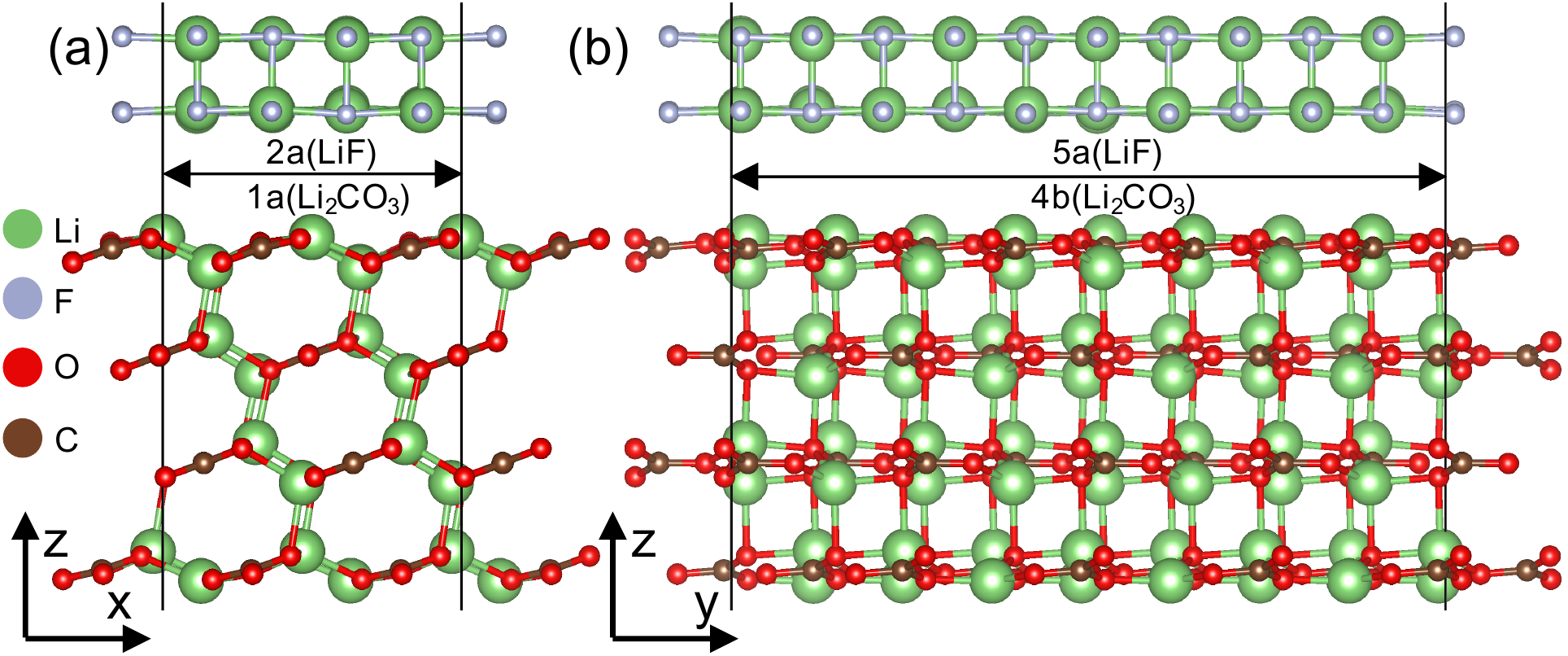}
    \caption{Geometry of the coherent interface between (001) \ce{LiF} surface and (001) \ce{Li2CO3} surface used in DFT calculations. (a) Front and (b) side views along with the axes orientation. The black line is the  boundary of the interfacial unit cell used in the simulations with the $z$ axis is perpendicular to the interface. The coherent interface was generated by using a (2,5,1) supercell of  LiF and (1,4,2) supercell of \ce{Li2CO3} as shown, resulting in a coherency strain of less than 3 \%.}
    \label{fig:interf}
\end{figure}

The interfacial distance between the two solids was obtained through energy minimization in two steps: first, single point energy calculations at different values of the distances were performed. This procedure gave a value of $\sim3.34$ {\AA}  for the distance between the closest Li layers of LiF and \ce{Li2CO3}. Finally, using this distance we relaxed the lattice parameters (in the plane of the interface) and the ionic positions using convergence criteria for forces $<0.001$ Ry/{\AA} and energy $<0.0001$ Ry. The relaxed interface geometry is shown in Fig. ~\ref{fig:interf}. For molecular dynamics calculations, only one repetition along z-axis (perpendicular to the interface) for both unit cells was used. The Li-O bonds on an average become shorter at the surface of \ce{Li2CO3} compared to the bulk on relaxation.

\begin{figure}
    \centering
    \includegraphics[scale=0.8]{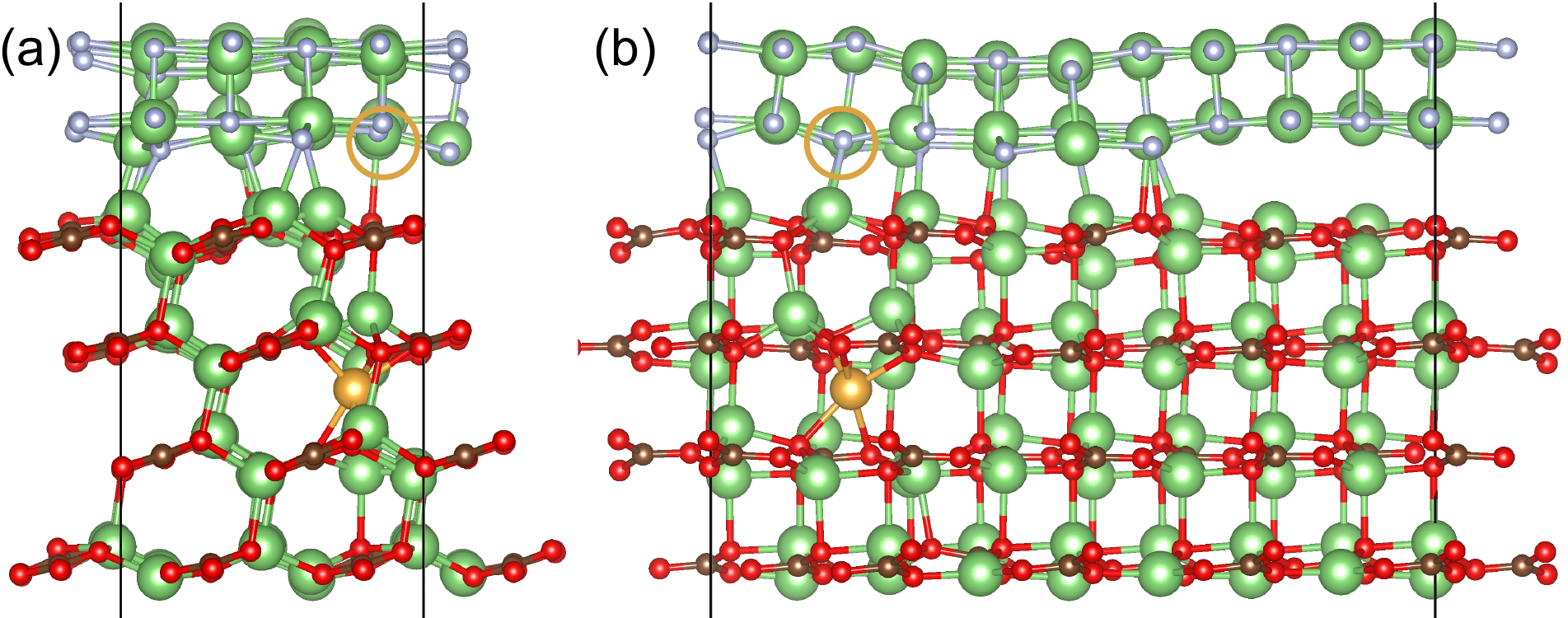}
    \caption{(a) Front and (b) side views of defective structure of the interface between LiF and \ce{Li2CO3}. A Li ion from LiF has moved to \ce{Li2CO3} creating an interstitial defect in \ce{Li2CO3} (yellow atom) and a Li vacancy in LiF (circled in yellow).}
    \label{fig:def_int}
\end{figure}

The interface between LiF and \ce{Li2CO3} in the SEI consists of undercoordinated atoms and hence, may be expected to host defects. ~\citeauthor{Zhang2016synergetic}~\cite{Zhang2016synergetic} proposed the following defect exchange reaction at the interface between LiF and \ce{Li2CO3} for the enhancement of Li ion conductivity of \ce{Li2CO3}:\\
\centerline{\ce{Li_{Li}(LiF) <=> Li_i^. (Li2CO3) + V_{Li}^'(LiF)}}
Based on defect formation energies and chemical potentials of interstitial Li and Li vacancy in bulk LiF and \ce{Li2CO3}, they obtained this as the only favorable reaction (-0.95 eV net energy change). The explicit treatment of the interface allows us to study the initial stage of this reaction when a Li atom migrates from LiF to \ce{Li2CO3}, forming an interstitial defect in \ce{Li2CO3} and Li vacancy in LiF. This state is shown in Fig.~\ref{fig:def_int}. The Li vacancy is circled in yellow and the Li interstitial is shown as a  yellow atom in \ce{Li2CO3}. The interstitial Li atom relaxes to a configuration with five Li-O bonds similar to lowest energy site II in Ref.~\citenum{Shi2012direct}. 
The energy required to move a Li ion from LiF to \ce{Li2CO3}, hence generating a Li vacancy in LiF and an interstitial Li in \ce{Li2CO3} can be calculated as:
\begin{equation}
    \Delta E = E_{\text{def}} - E_{\text{pris}}
\end{equation}
where $E_{\text{def}}$ is the energy of the defective system and $E_{\text{pris}}$ is the energy of the pristine system. Our interface calculations show that the energy required to move a Li ion from LiF to \ce{Li2CO3} in the configuration shown in Fig.~\ref{fig:def_int} is 1.05 eV. This is an intermediate stage of movement of the defects into the bulk of these solids. The distance between the Li vacancy and interstitial in this Frenkel pair is approximately 7 \AA. A smaller separation distance may result in a lower formation energy due to electrostatic attraction, however, we were unable to relax the Frenkel pair located any closer due to the relaxation to the pristine configuration. Our result shows that although the defect exchange reaction is favorable in the bulk, there is a significant barrier in the process as the Li ion moves from the interface towards the bulk of \ce{Li2CO3}.

\subsection{Electronic Properties}
\begin{figure}[htbp]
    \centering
    \includegraphics[scale=0.8]{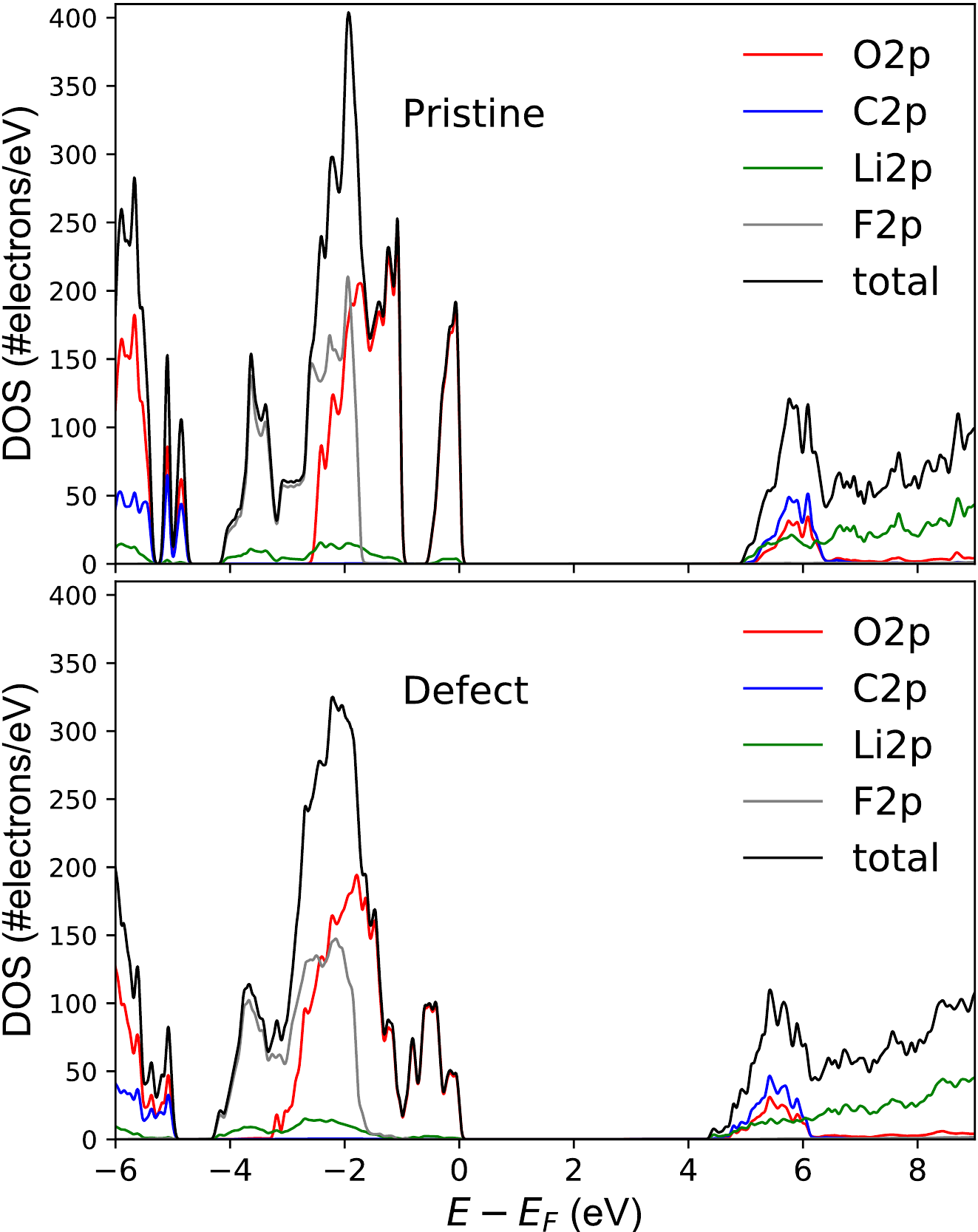}
    \caption{Density of states for the pristine and defective structure of the interface between LiF and \ce{Li2CO3}. The Fermi level ($E_F$) is set to VBM. Notably, no defect states in the band gap are found. The O 2p orbital DOS becomes diffuse due to distortion caused by the interstitial Li in \ce{Li2CO3}. The Li intersitial defect causes a reduction in the PBE band gap by 0.49 eV.}
    \label{fig:dos}
\end{figure}

Recently, electronic conductivity of interface between electrode and electrolyte has been proposed as the cause of dendrite growth in solid electrolytes~\cite{Han2019high, Tian2019interfacial}. The enhanced dendrite growth at higher temperatures observed by ~\citeauthor{Han2019high}~\cite{Han2019high} points to an important role of electronic leakage through interfaces in promoting dendrite growth. We performed DFT calculations to obtain the electronic density of states (DOS) of the pristine and defective interfacial structures. Figure~\ref{fig:dos} compares the electronic DOS for the pristine structure and the one that contains an interstitial Li in \ce{Li2CO3} from \ce{LiF}. We find that the valence band minimum (VBM) is primarily composed of O 2p states from \ce{Li2CO3} in both cases.  The interstitial Li causes distortion of the O atoms in \ce{Li2CO3} and creates a more diffuse O 2p DOS near the VBM. This is in contrast to the results for a Li vacancy in \ce{Li2CO3} which results in formation of localized holes near the VBM~\cite{Chen2011electrical}. The electron densities for the two cases are compared in Fig. S3 of the Supporting Information. The PBE band gap reduces from 4.97 eV in the pristine structure to 4.48 eV for the defective structure.  Although this is a lower bound on the band gap as is expected from semilocal DFT, we do not expect a drastic change in the occupied VBM states by incorporating a Hubbard U for O~\cite{Gajdo2005CO}. Fig. S2 show the contribution of the interstitial Li states to the DOS mainly at the conduction band minimum, causing a reduction of the band gap in the defective structure. The electronic DOS results show that electronic transport through the SEI even in the presence of these heterointerfaces is very small.

\subsection{Li Hopping Barrier}
\begin{figure}
    \centering
    \includegraphics[width=.75\textwidth]{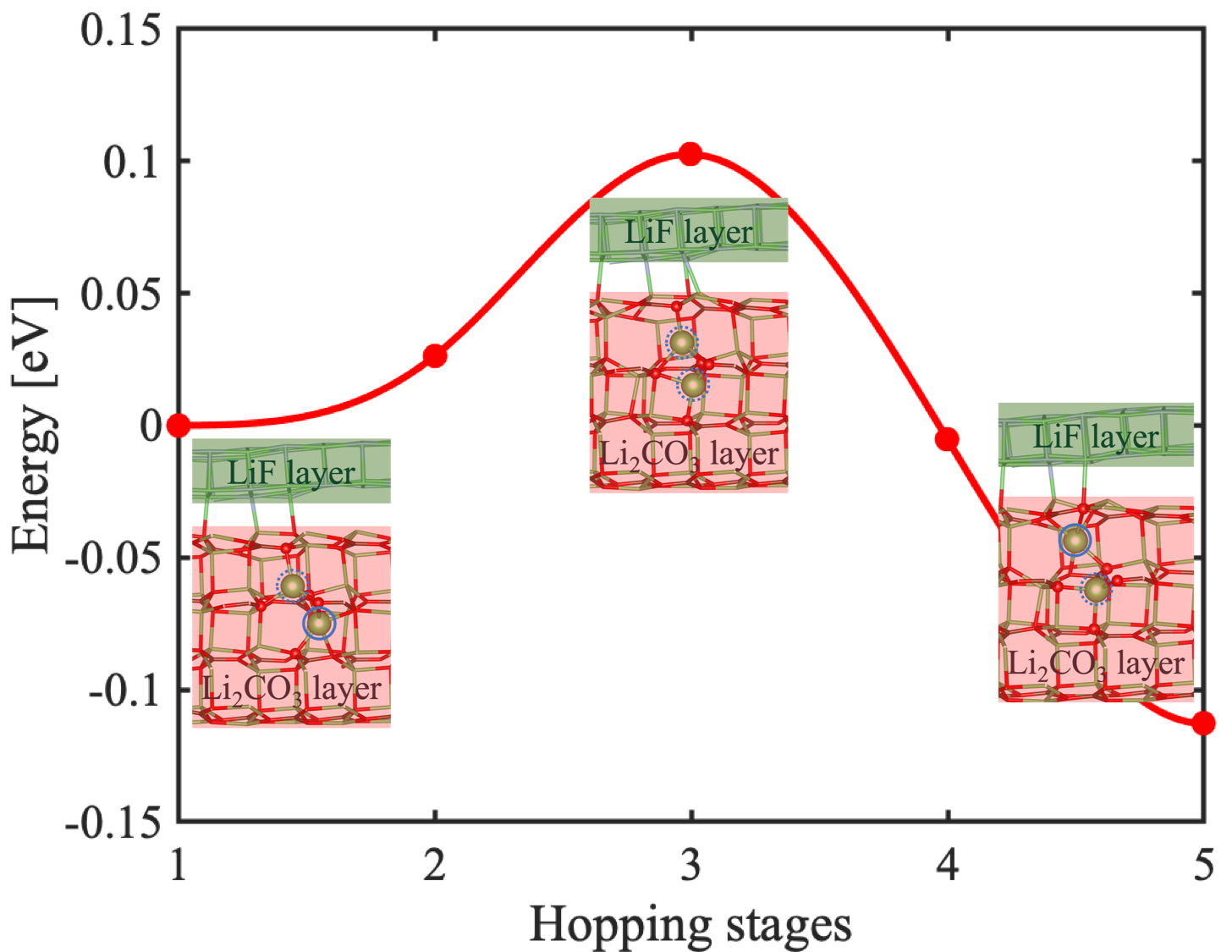}
    \caption{Energetics of minimum energy pathway of interstitial Li motion in \ce{Li2CO3} layer. Interstitial Li atoms are circled by a continuous blue line, while lattice Li atoms that partake in the motion are circled by a dashed blue line. The activation energy of Li hopping is reduced from $0.3$ eV in bulk \ce{Li2CO3} to $0.10$ or $0.22$ eV in the interfacial structure considered, showing that the interface assists in Li diffusion.}
    \label{fig:MEP}
\end{figure}

Various mechanisms of Li hopping in SEI components have been proposed. ~\citeauthor{Chen2011electrical}~\cite{Chen2011electrical} performed a study of Li hopping mechanisms in the SEI components LiF, \ce{Li2O} and \ce{Li2CO3} in the presence of Li vacancies. However, an exhaustive study of the possible point defects and transport mechanisms in \ce{Li2CO3} found that Li interstitials have the lowest formation energy and are responsible for transport, especially at the anode side~\cite{Shi2013defect}. The most favorable pathway for ionic motion in this system is by the \textit{knock-off} or ``interstitialcy'' mechanism, a form of indirect motion in which an interstitial atom displaces an atom in a lattice site into a neighboring interstitial site, and occupies the now vacant lattice site. The authors argue that, by maintaining a high cation-anion coordination number through all stages of motion, the knock-off mechanism lowers the barrier associated with ion transport when compared to a direct interstitial pathway. 
Here, we study the effect of the interface between LiF and \ce{Li2CO3} on Li hopping in the latter. Fig.~\ref{fig:MEP} shows the energy landscape of the Li interstitial in  \ce{Li2CO3} near its interface with \ce{LiF} based on this knock-off Li transport. Note that the first and the final hopping stages are not equivalent: given that the intended Li motion happens along the [011] direction, the initial interstitial atom has to be placed further from the interface than in the final configuration, as shown in the insets of Fig.~\ref{fig:MEP}. The activation energy required to move an atom from one of these interstitial sites to another is thus dependent on the original site, and can be either $0.10$ or $0.22$ eV. Therefore, the average hopping barrier associated with a  net motion pathway parallel to the LiF-\ce{Li2CO3} interface is of $\sim0.16$ eV. The same knock-off mechanism in bulk \ce{Li2CO3} has an activation energy of 0.3 eV~\cite{Shi2012direct}, further reinforcing the hypothesis that Li transport is enhanced at the interface of LiF and \ce{Li2CO3}. The activation energy is also much lower than that for Li hopping at a surface of LiF (0.34 eV)~\cite{Fu2020universal}.

\subsection{Dynamics}
A simulation of the finite temperature dynamics can provide insights into ion transport in the SEI, which is the rate determining step for many electrolyte-electrolyte systems~\cite{balbuena2004lithium}. The multiphase nature of the SEI necessitates an investigation of Li ion transport incorporating the effect of phase boundaries. To understand the effect of interface on Li ion conduction, we performed molecular dynamics simulations of the LiF-\ce{Li2CO3} interfacial structure. The Li radial distribution functions at different temperatures obtained from the simulations are plotted in the Supporting Information. Previously, ~\citeauthor{Mizusaki1992lithium}~\cite{Mizusaki1992lithium} studied the conductivity of \ce{Li2CO3} as a function of dopant concentration using impedance spectrometry. They found that the conductivity for pure \ce{Li2CO3} along ab plane was much higher than perpendicular to it and  largely due to interstitial Li ions. Fitting of their separate plots of ionic conductivity along and perpendicular to that plane to Arrhenius equation give activation energies of 0.49 and 0.62 eV respectively. The higher conductivity along ab plane may be caused by the growth along (002) plane of the crystal. \citeauthor{Shi2012direct}~\cite{Shi2012direct} studied the atomistic mechanisms of Li ion transport in crystalline \ce{Li2CO3}. They proposed a new mechanism  by which Li ions can hop through a series of steps with low activation energy. The activation energy of the proposed \emph{knock-off} was found to be smaller (0.3 eV) compared to the direct hopping mechanism (0.54 eV). ~\citeauthor{Benitez2017ion}~\cite{Benitez2017ion} studied the mechanism of Li ion conduction in the SEI components LiF, \ce{Li2O} and \ce{Li2CO3} using molecular dynamics based on force fields. They also inferred the \emph{knock-off} mechanism of Li ion transport in \ce{Li2CO3}. ~\citeauthor{Ramasubramanian2019lithium}~\cite{Ramasubramanian2019lithium} studied the conduction properties of grain boundaries between the components LiF and \ce{Li2O} in the SEI. They ignored \ce{Li2CO3} as a component for grain boundary conduction analysis due to the possibility of reduction in the presence of Li. We, however, consider it an important constituent for efficient Li ion transport due to strong recent evidence suggesting its presence in the SEI~\cite{Mozhzhukhina2020direct}.

\begin{figure}
    \centering
    \begin{subfigure}{.45\textwidth}
    \centering
    \includegraphics[scale=0.5]{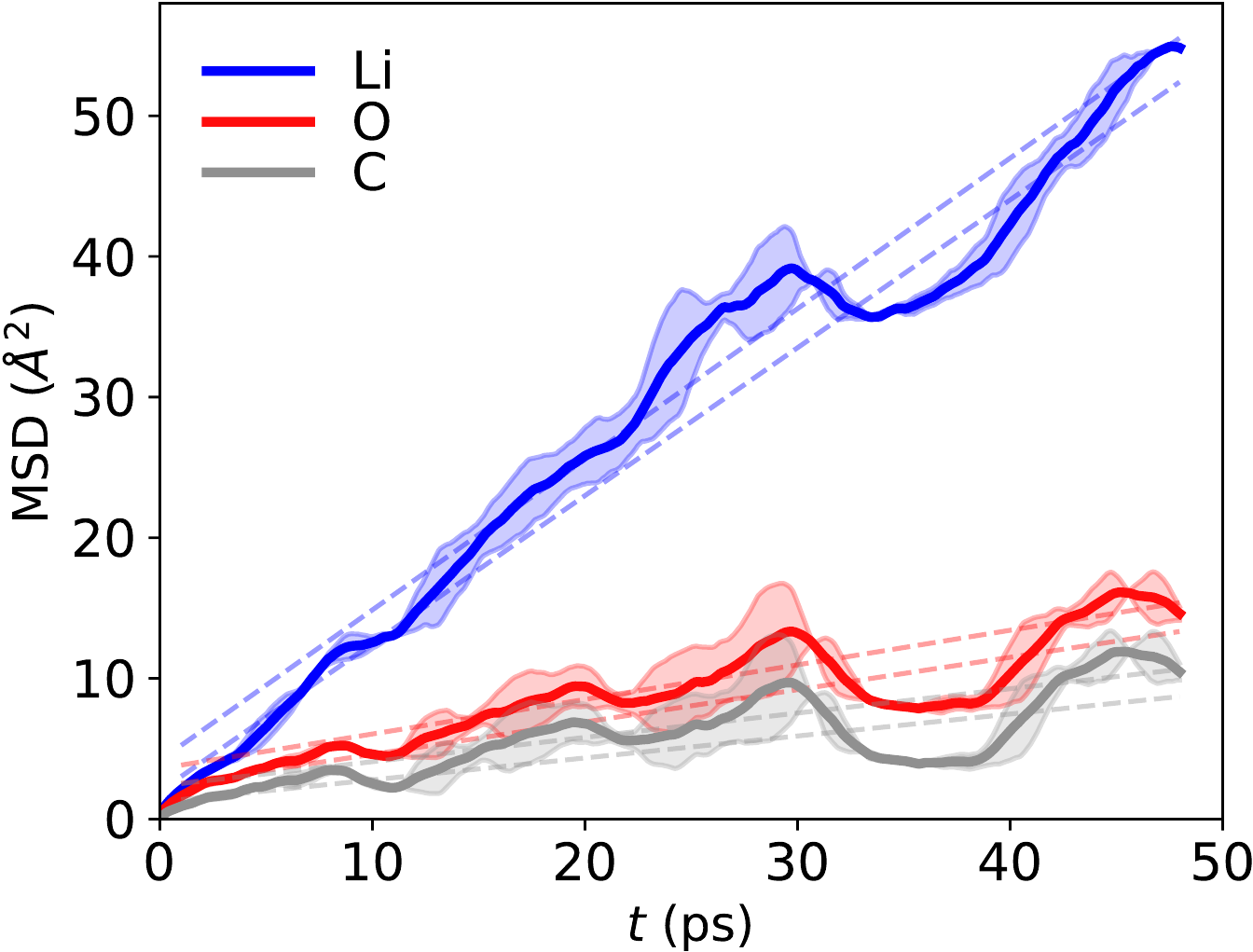}
    \caption{\label{fig:msdLi}}
    \end{subfigure}
    \begin{subfigure}{.45\textwidth}
    \centering
    \includegraphics[scale=0.5]{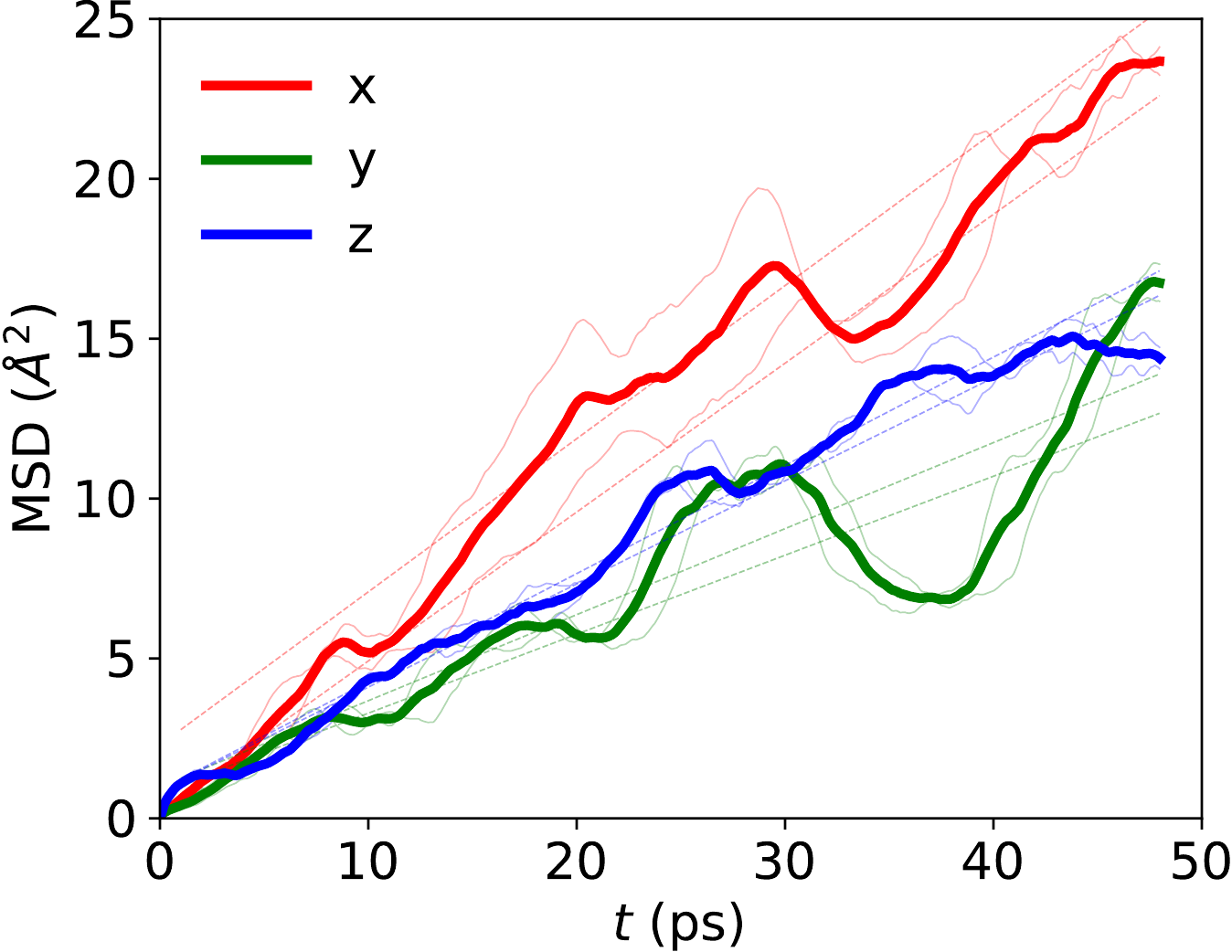}
    \caption{\label{fig:msdLixyz}}
    \end{subfigure}
    \caption{(a) MSD for Li, O and C ions in \ce{Li2CO3} interfaced with LiF at 800 K. Linear fits with time are shown by dashed lines. The diffusion coefficient of Li ions is $1.77\times 10^{-5}$ cm$^2$/s. (b) Diagonal components of the tensor $\rtten{T}_{ij} = \Delta R_i \Delta R_j $ for Li ions in \ce{Li2CO3} at 800 K.}
\end{figure}

\begin{figure}
    \centering
    \includegraphics[scale=0.6]{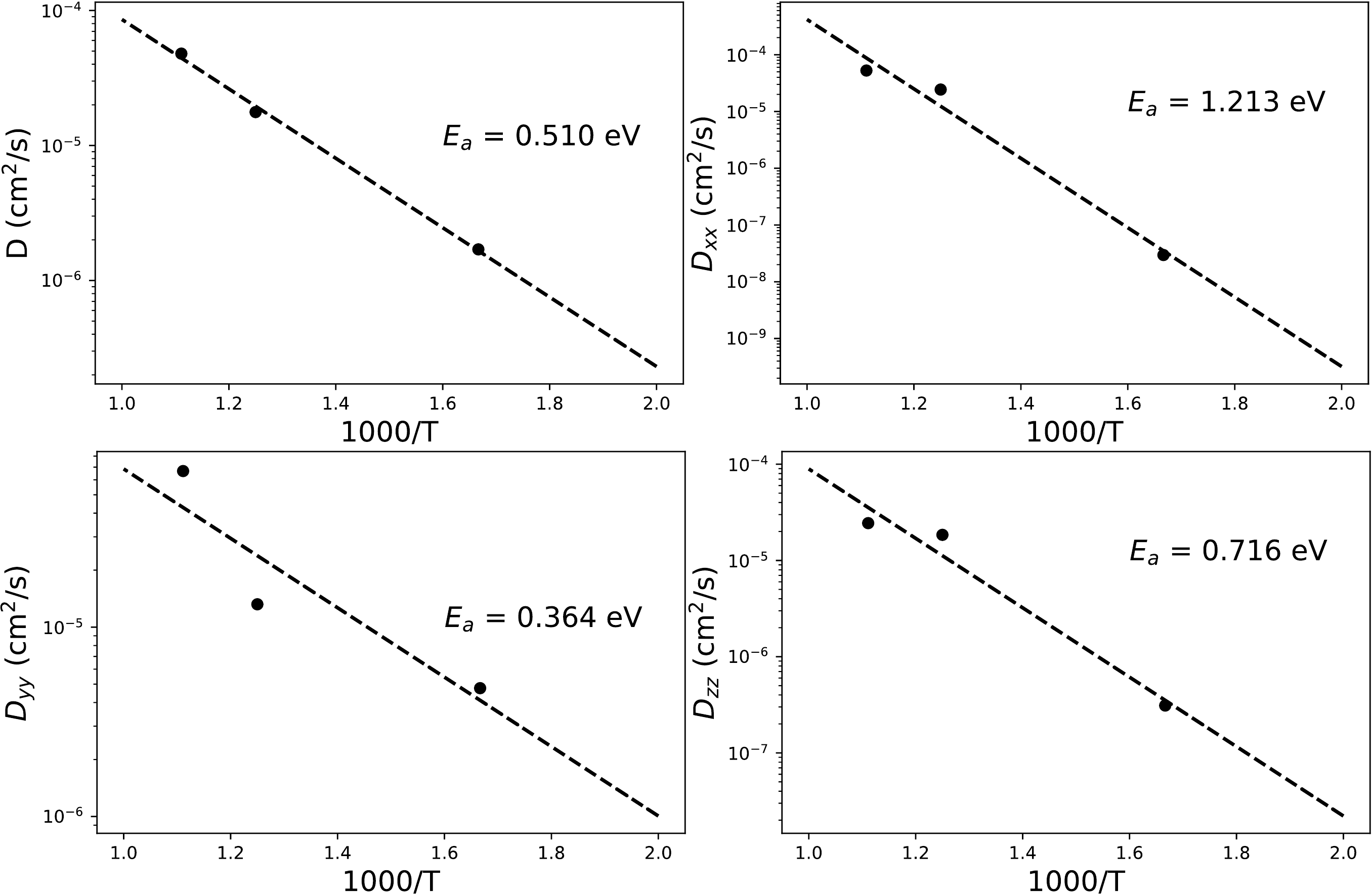}
    \caption{Variation of diffusion coefficient and the diagonal components of diffusion coefficient tensor $D_{xx}$, $D_{yy}$ and $D_{zz}$ with temperature $T$. The straight line represents an Arrhenius fit. The existence of fast diffusion pathways results in  a lower activation energy along the $y$ direction.}
    \label{fig:DvsT}
\end{figure}

\begin{figure}
    \centering
    \includegraphics[scale=0.7]{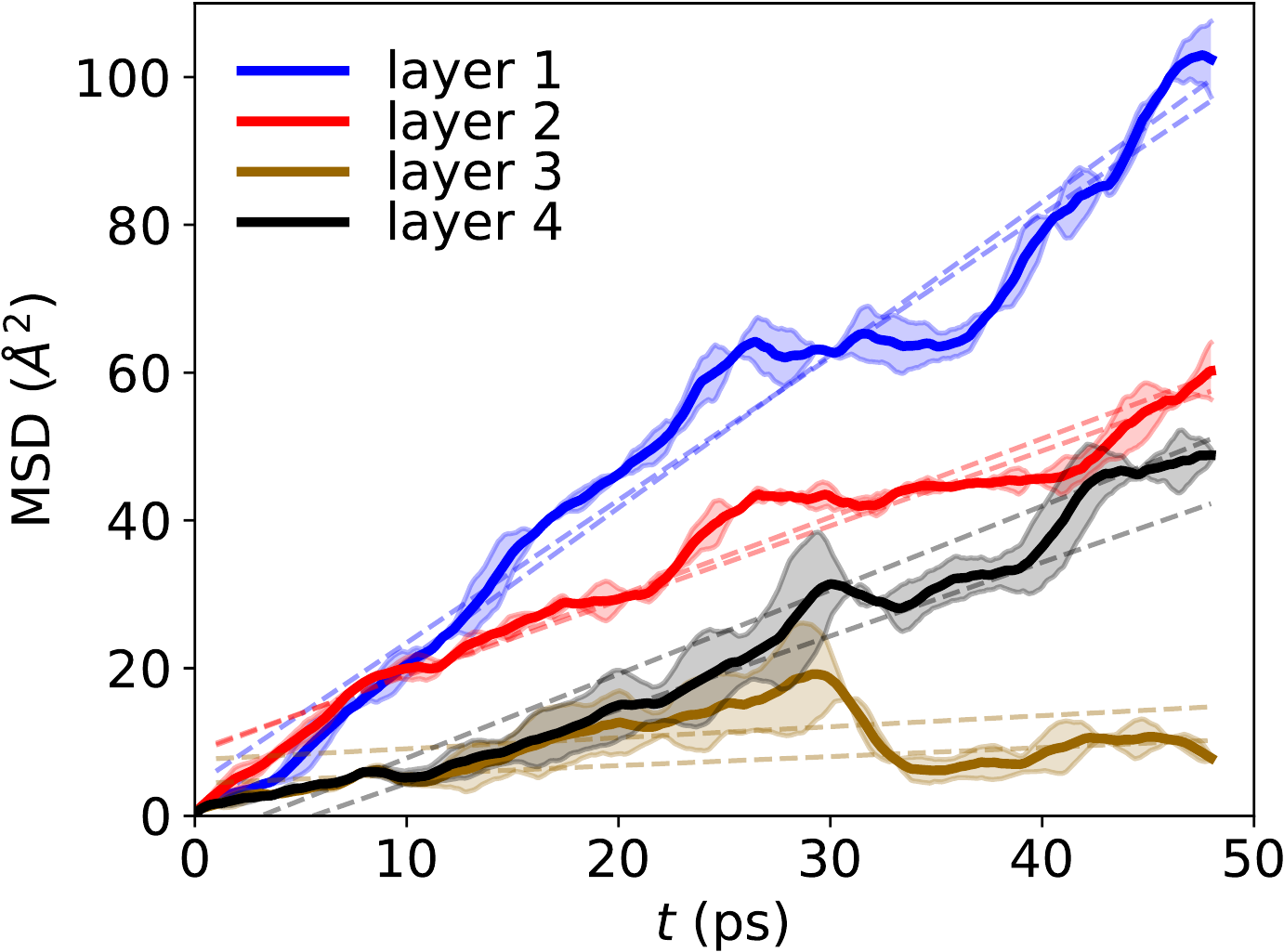}
    \caption{MSD for Li ions in \ce{Li2CO3} vs time calculated layerwise at 800 K. Layer 1 is closest to the interface while layer 4 is the farthest. The layers closer to the interface have the highest MSD. The higher value for layer 4 compared to layer 3 may be due to its exposure to vacuum.}
    \label{fig:msdLilayers}
\end{figure}

We quantify the Li ion conduction properties of the interface through the self-diffusion coefficient of Li ions in \ce{Li2CO3}. The analysis of the Car-Parrinello molecular dynamics trajectories is done using the Suite for Analysis of MOlecular Simulations (SAMOS) package~\cite{samosGihub}. 
The diffusion coefficient $D$ can be calculated using a linear fit of the mean squared displacement (MSD) of Li ions as a function of time $t$:
\begin{equation}
    D = \frac{1}{2dt}\langle \Delta R(t)^2 \rangle
\end{equation}
where $d$ is the dimension ($d=3$ for a three-dimensional simulation) and $\Delta R$ is the ionic displacement. Fig.~\ref{fig:msdLi} shows the MSD for Li, O and C ions in \ce{Li2CO3} at 800 K and Fig. \ref{fig:msdLixyz} shows the diagonal components of the tensor $\rtten{T}_{ij} = \Delta R_i \Delta R_j $ for Li ions in \ce{Li2CO3}. 
Here, the indices $i$, $j$ stand for $\{x,y,z\}$ directions along which the displacement is calculated. The diffusive nature of Li ions is manifested by the linear increase of MSD with time. The same molecular dynamics calculations were performed at 600 and 900 K to obtain the activation energies for Li hopping. Fig.~\ref{fig:DvsT} shows the variation of diffusion coefficient with temperature and the fit to the Arrhenius relationship. 
Although the mean squared deviation along the $x$ direction seems to be the highest at 800 K, the Li hopping along this direction also has the highest activation energy resulting in much reduced number of hops at room temperature. At 300 K, we obtain a Li ion diffusivity in \ce{Li2CO3} equal to $7.6\times 10^{-11}$ cm$^2$/s. This is an order of magnitude higher than LiF/LiF and \ce{Li2O}/\ce{Li2O} grain boundaries and five times lower than LiF/\ce{Li2O} phase boundary~\cite{Ramasubramanian2019lithium}. The differences may be due to the material or the method used to calculate diffusivity. Our interfacial structure enables us to study the layer dependent diffusivity of Li ion \ce{Li2CO3}, and hence incorporate the effect of the interface ion conduction. The two Li layers closest to the interface have the highest Li ion diffusion coefficients as seen in Fig.~\ref{fig:msdLilayers}. The diffusion coefficient in \ce{Li2CO3} is about four orders of magnitude lower than values for a typical liquid electrolyte ($\sim 10^{-6}$cm$^2$/s)~\cite{Kitz2020operando}.

The overall activation energy for Li diffusion in \ce{Li2CO3} is 0.51 eV. An overall activation energy is obtained by averaging over the different pathways for Li hopping and hence is not solely determined by the least activation energy pathway. Since the magnitude of mean squared displacement at 800 K is significant along $x$ direction with a higher activation energy, the overall activation energy includes contributions along that pathway and is much higher than that obtained by considering only the knockoff mechanism. The activation energy for diffusion along $y$ direction is 0.36 eV and is the lowest due to low activation energy pathways along this direction. Note that this value of self-diffusion is on par with the activation barrier of  0.31 eV~\cite{Shi2012direct} for the optimal ionic motion pathway in bulk \ce{Li2CO3} and close to 0.28 eV obtained by ~\citeauthor{Iddir2010Li}~\cite{Iddir2010Li} considering motion along [010] open channels, further indicating that, on average, interfacial Li diffusion (whether mediated by interstitial atoms or otherwise) is faster than in the bulk. Solid-state NMR spectroscopy measurements revealed an activation energy for Li ion hopping in bulk \ce{Li2CO3} in the range 0.78-1.34 eV~\cite{Dunstan2015ion}. Recent electrochemical impedance spectroscopy measurements suggested an activation energy of 0.42 eV and 0.58 eV for ionic conduction in LiF and \ce{Li2O} SEI respectively~\cite{Guo2020Li2O}. Based on the considered LiF-\ce{Li2CO3} structure (Fig.~\ref{fig:interf}), we find that it is beneficial to have the plane of the lattice matched interface between LiF and \ce{Li2CO3} perpendicular to the electrode-electrolyte interface to benefit from fast ion conduction along [010] direction. To further understand the mobility of Li ions in different layers of \ce{Li2CO3}, we generated probability distributions of Li ions originating from \ce{Li2CO3} and LiF along the axis perpendicular to the interface ($z$) at 600 and 800 K as shown in Fig. S6. We observe the existence of distinct peaks in the probability distribution corresponding to the different layers in LiF and \ce{Li2CO3}. There is a significant probability of Li ions from \ce{Li2CO3} to migrate to LiF and vice versa at 800 and 900 K (not shown) but not at lower temperatures. However, the presence of defects might promote Li ion exchange between the two materials even at lower temperatures.


\section{Summary \& Conclusions}
We have probed the interface of canonical solid electrolyte interphase (SEI) components, LiF and \ce{Li2CO3}.  We studied the effects of the interface on electronic transport, Li ion conduction and defect formation in \ce{Li2CO3}. Our results show that the generation of defects at the interface (Frenkel pair) requires significant energy compared to the analysis based on bulk properties. The molecular dynamics calculations reveal the Li ions starting closest to the LiF layer have the highest MSDs and hence the highest diffusion coefficients. The temperature dependence of the diffusion coefficients gives insights into the activation energies for hopping along different directions. The lowest activation energy of 0.36 eV is obtained along the \hkl[010] direction in agreement with our results on the existence of a low activation energy pathway for Li migration in this direction in \ce{Li2CO3}. By showing the stark difference between interfacial and bulk ionic transport phenomena, our work demonstrates that interfaces are of utmost relevance when trying to understand the SEI, ``one of the most important, yet least understood,''~\cite{winter2009solid} battery component. We note that our results rely on a crystalline structure of the multiphase SEI. Amorphous regions generated in multiphase SEI might disrupt Li ion diffusion pathways and result in a lower ionic conductivity than the single-phase SEI~\cite{Guo2020Li2O}. A highly heterogeneous SEI might also result in varying current densities in different regions of the electrode-electrolyte interface, potentially leading to dendrite growth in Li metal batteries. A careful design of the SEI needs to be pursued to harvest the beneficial effects of interfaces in multiphase systems.

\begin{acknowledgement}

We thank A. Khetan for a critical reading of the manuscript. Z.A. and V.Venturi acknowledge support from the Advanced Research Projects Agency Energy (ARPA-E) under Grant DE-AR0000774. This work used the Extreme Science and Engineering Discovery Environment (XSEDE), which is supported by National Science Foundation grant number ACI-1548562~\cite{towns2014xsede}. Specifically, it used the Bridges system, which is supported by NSF award number ACI-1445606, at the Pittsburgh Supercomputing Center (PSC)~\cite{nystrom2015psc}.

\end{acknowledgement}

\begin{suppinfo}
The following files are available free of charge.
\begin{itemize}
\item suppinfo.pdf: Details of electronic density of states for the pristine and defective structures 
\end{itemize}
The code used for the analysis of molecular dynamics simulations is available on GitHub:\\ \href{https://github.com/ahzeeshan/samos}{https://github.com/ahzeeshan/samos}. Relaxed configurations of the pristine and defective structures of the interface in the form of cif files are available from the GitHub repository \href{https://github.com/BattModels/SEI-interface}{https://github.com/BattModels/SEI-interface}.

\end{suppinfo}

\bibliography{refs}
\includepdf[pages=1-6]{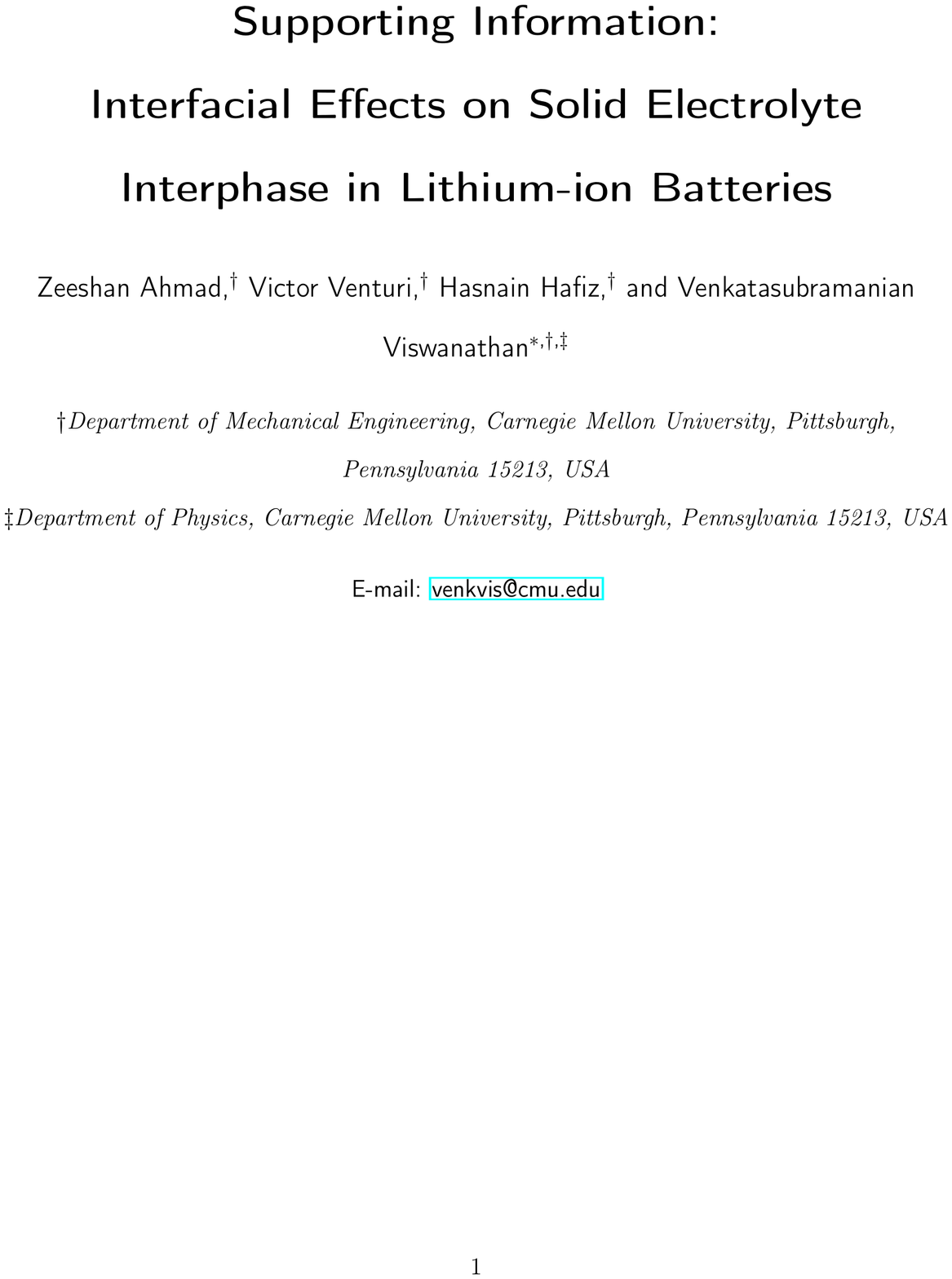}

\end{document}